\documentclass[twocolumn,pr,showpacs,amsfonts,amsmath,assymb,eufrak,superscriptaddress]{revtex4-1}
\usepackage{amsmath,amsfonts,amssymb,mathtools,upgreek}
\usepackage{epsfig}

\def\eq#1\en{\begin{equation}#1\end{equation}}  
\def\eqa#1\ena{\begin{align}#1\end{align}}
\def\eqg#1\eng{\begin{gather}#1\end{gather}}
\newcommand{\lb}[1]{\label{e:#1}}
\newcommand{\rlb}[1]{\eqref{e:#1}} 
\newcommand{\nl}{\notag\\}

\newcommand{\sbkt}[1]{\langle#1\rangle}

\newcommand{\sumtwo}[2]%
{\mathop{\sum_{#1}}_{#2}}
\newcommand{\sumthree}[3]%
{\mathop{\mathop{\sum_{#1}}_{#2}}_{#3}}
\newcommand{\sumfour}[4]%
{\mathop{\mathop{\mathop{\sum_{#1}}_{#2}}_{#3}}_{#4}} 
\newcommand{\prodtwo}[2]%
{\mathop{\prod_{#1}}_{#2}}
\newcommand{\mintwo}[2]%
{\mathop{\min_{#1}}_{#2}}
\newcommand{\maxtwo}[2]%
{\mathop{\max_{#1}}_{#2}}
\newcommand{\maxthree}[3]%
{\mathop{\mathop{\max_{#1}}_{#2}}_{#3}}
\newcommand{\limtwo}[2]%
{\mathop{\lim_{#1}}_{#2}}
\newcommand{\suptwo}[2]%
{\mathop{\sup_{#1}}_{#2}}
\newcommand{\supthree}[3]%
{\mathop{\mathop{\sup_{#1}}_{#2}}_{#3}}
\newcommand{\supfour}[4]%
{\mathop{\mathop{\mathop{\sup_{#1}}_{#2}}_{#3}}_{#4}} 
\newcommand{\inftwo}[2]%
{\mathop{\inf_{#1}}_{#2}}
\newcommand{\infthree}[3]%
{\mathop{\mathop{\inf_{#1}}_{#2}}_{#3}}
\newcommand{\inffour}[4]%
{\mathop{\mathop{\mathop{\inf_{#1}}_{#2}}_{#3}}_{#4}} 
\newcommand\calB{{\cal B}}

\newcommand\calE{{\cal E}}

\newcommand\calI{{\cal I}}

\newcommand\calK{{\cal K}}
\newcommand\calL{{\cal L}}

\newcommand\calN{{\cal N}}

\newcommand{\bsn}{\boldsymbol{n}}

\newcommand{\bbC}{\mathbb{C}}

\newcommand{\up}{\uparrow}

\newcommand{\Di}{\mathit{\Delta}}
\newcommand{\La}{\Lambda}
\newcommand{\qedm}{\rule{1.5mm}{3mm}}
\newcommand{\ha}{\hat{a}}
\newcommand{\had}{\hat{a}^\dagger}
\newcommand{\hb}{\hat{b}}
\newcommand{\hbd}{\hat{b}^\dagger}
\newcommand{\hd}{\hat{d}}
\newcommand{\hdd}{\hat{d}^\dagger}
\newcommand{\hn}{\hat{n}}
\newcommand{\hH}{\hat{H}}
\newcommand{\Hhop}{\hat{H}_{\rm hop}}
\newcommand{\Hint}{\hat{H}_{\rm int}}

\newcommand{\ket}[1]{|#1\rangle}
\newcommand{\vac}{\ket{\Phi_{\rm vac}}}
\newcommand{\vacb}{\langle\Phi_{\rm vac}|}
\newcommand{\GS}{\ket{\Phi_{\rm GS}}}
\newcommand{\GSb}{\langle\Phi_{\rm GS}|}
\newcommand{\Phin}{\ket{\Phi_{\bsn}}}
\newcommand{\per}{\uppi}

\newcommand{\GSp}{\ket{\Phi_{\rm GS}^+}}
\newcommand{\GSpb}{\langle\Phi_{\rm GS}^+|}
\newcommand{\GSpbn}{\langle\Phi_{\rm GS}^+}

\newcommand{\bktgs}[1]{\sbkt{#1}_{\rm GS}}

\newcommand{\DI}{\partial\calI}
\newcommand{\betac}{\beta_{\rm c}}
\newcommand{\tbeta}{\tilde{\beta}}

\newcommand{\hD}{\hat{D}_0}
\newcommand{\hDD}{\hat{D}^\dagger_0}

\newcommand{\EGS}{E^{\rm GS}}

\newcommand{\prop}[1]{\medskip{\em #1}\/.---\ }
\begin{document}
\title{Mott Insulator-like Bose-Einstein Condensation \\in a Tight-Binding System of Interacting Bosons with a Flat Band}

\author{Hosho Katsura}
\affiliation{Department of Physics, Graduate School of Science, The University of Tokyo, 7-3-1 Hongo, Tokyo 113-0033, Japan}
\affiliation{Institute for Physics of Intelligence, The University of Tokyo, 7-3-1 Hongo, Tokyo 113-0033, Japan}
\affiliation{Trans-scale Quantum Science Institute, The University of Tokyo, 7-3-1, Hongo, Tokyo 113-0033, Japan}

\author{Naoki Kawashima}
\affiliation{The Institute for Solid State Physics, The University of Tokyo, Kashiwa, Chiba 277-8581, Japan}

\author{Satoshi Morita}
\affiliation{The Institute for Solid State Physics, The University of Tokyo, Kashiwa, Chiba 277-8581, Japan}

\author{Akinori Tanaka}
\affiliation{Department of General Education, National Institute of Technology, Ariake College, Omuta, Fukuoka 836-8585, Japan}

\author{Hal Tasaki}
\affiliation{Department of Physics, Gakushuin University, Mejiro, Toshima-ku, Tokyo 171-8588, Japan}

\date{\today}

%%%%%%%%%%%%%%%%%
\begin{abstract}
We propose a new class of tight-binding systems of interacting bosons with a flat band, which are exactly solvable in the sense that one can explicitly write down the unique ground state.
The ground state is expressed in terms of local creation operators,
and apparently resembles that of a Mott insulator.
Based on an exact representation in terms of a classical loop-gas model, we conjecture that the ground state may exhibit quasi Bose-Einstein condensation (BEC) or genuine BEC in dimensions two and three or higher, respectively, still keeping Mott insulator-like character.
Our Monte Carlo simulation of the loop-gas model strongly supports this conjecture, i.e.,
the ground state undergoes a Kosterlitz-Thouless transition and exhibits quasi BEC in two dimensions.
(There is a 22.5 minutes video in which the main results of the paper are described.  See \url{https://youtu.be/ef9x11C1cVg})
\end{abstract}


\maketitle
%%%%%%%%%%%%%%%%%%%%%%%%%%%%%%%%%%%%%
%%%%%%%%%%%%%%%%%%%%%%%%%%%%%%%%%%%%%

\section{Introduction}%
Tight-binding models of interacting particles with a flat band, i.e., a set of highly degenerate single-particle energy eigenstates, have been studied intensively over the decades.
Flat-band systems do not only serve as idealized models of materials with a narrow band, but also provide a theoretical playground for investigating various collective phenomena arising from the interplay between particle motion and interactions.
This is because the effect of interactions is magnified due to the flatness of the band.
Such an approach was  fruitful in the study of the origin of ferrimagnetism  \cite{Lieb89} and ferromagnetism \cite{Mielke91b,Mielke92,Tasaki92e,MielkeTasaki1993} in the Hubbard model.
See \cite{TasakiBook} for a review.
For the formation of a Wigner crystal and the effect of the change in the density in bosonic systems with a flat band, see \cite{HuberAltman,TakayoshiKatsuraWatanabeAoki,TovmasyanvanNieuwenburgHuber,Mielke2018,FronkMielke2020}. 
The recent proposal that the nearly flat band in twisted bilayer graphene supports superconductivity is  intriguing \cite{graphene}.

In this paper we propose a  new class of tight-binding systems of interacting bosons with a flat lowest band.
The model is based on the construction in \cite{Tasaki1998B} of flat band Hubbard models, and can be regarded as a spinless version of the models studied in \cite{YangNakanoKatsura}.
We write down the ground state of the model explicitly as in \rlb{GS}, and prove that it is the unique ground state.
The expression suggests that the ground state is essentially different from that of a non-interacting system, and resembles that of a Mott insulator.

In spite of the simple expression, the property of the ground state is nontrivial and rich.
By examining representations of the norm and the correlation functions in terms of a classical loop-gas model, we conjecture that the ground states may exhibit quasi off-diagonal long-range order (ODLRO) in two dimensions, and genuine ODLRO in three or higher dimensions.
We present some results of Monte Carlo simulation of the loop-gas model, which strongly indicates that the two-dimensional model undergoes a Kosterlitz-Thouless (KT) transition and exhibits quasi ODLRO in its ground states.

We note that a class of states (without parent Hamiltonians) very similar to ours was proposed and examined in \cite{Kimchi}.
It was found that these states do not exhibit off-diagonal (quasi) long-range order.

It was found in a two-component system of bosons that one component may exhibit Bose-Einstein condensation (BEC) while the other is in the Mott insulating state \cite{ChenWu}.
This is different from our Mott insulator-like ground state \rlb{GS}, which consists only of the states created by the $\hbd_u$ operators.
Note also that our ground state in the BEC phase, although solid-like, is not a supersolid since there is no spontaneous breakdown of translation symmetry \cite{KimChan2004,OhgoeSuzukiKawashima2012,SuzukiKoga2014}.
It is a challenging problem to design a similar exactly solvable model that has supersolid ground states.

We stress that our ground states maintain Mott-insulating-like nature even when they exhibit (quasi) ODLRO.
This is most clearly seen in the anomalously small particle number fluctuation observed in a specific setting.
See section~\ref{s:discussion}, in particular, Figure~\ref{f:coupled}.
This does not mean, however, that the ground states describe genuine Mott insulators.  See section~\ref{s:OPN}.

It would be exciting if our exactly solvable model provides an example of a novel exotic phase of matter where Mott-insulator like nature and (quasi) ODLRO coexists.
Although we are not able to give a definite conclusion at the moment, we find it rather likely that (unfortunately) our Mott-insulator like condensate is smoothly connected to ordinary Bose-Einstein condensate realized, e.g., in non-interacting systems.
See section~\ref{s:discussion}.

We nevertheless believe that it is of essential importance that an entirely new class of exactly solvable models of strongly interacting bosons has been discovered.
We hope that the present work opens a new direction in the research of quantum many-body systems.

\section{The model and the exact ground state}%
\begin{figure}
\centerline{\epsfig{file=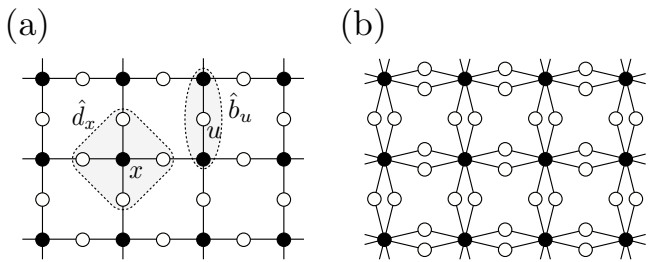,width=8.5cm}}
\caption[dummy]{
(a)~The lattice $\La$ in the case $(\calE,\calB)$ is the square lattice ($d=2$, $p=1$).
The black and white dots denote the sites in $\calE$ and $\calI$, respectively.
The states corresponding to the $\hb$ and $\hd$ operators are also shown.
(b)~The lattice $\La$ obtained from the square lattice in which neighboring sites are connected by two bonds ($d=2$, $p=2$).
}
\label{f:bd}
\end{figure}

\label{s:def}
Let $(\calE,\calB)$ be a finite lattice, where $\calE$ is the set of sites and $\calB$ is the set of bonds.
A bond is an unoriented segment that connects two distinct sites in $\calE$.
We allow multiple bonds to connect the same pair of sites.
At the center of each bond $b\in\calB$, we take a new site and denote it as $u_b$.
We let $\calI$ be the set of all sites $u_b$ with $b\in\calB$, and consider  the decorated lattice $\La=\calE\cup\calI$.
We typically choose $\calE$ to be the set of sites of the $d$-dimensional hypercubic lattice with periodic boundary conditions, and take $p$ distinct bonds connecting a pair of neighboring sites.
See Fig.~\ref{f:bd} for the resulting decorated lattices.

We shall define a tight-binding model of bosons on $\La$.
We denote by $\had_r$ and $\ha_r$ the creation and annihilation operators, respectively, of a boson at site $r\in\La$.
They satisfy the canonical commutation relations $[\ha_r,\ha_s]=0$ and $[\ha_r,\had_s]=\delta_{r,s}$ for any $r,s\in\La$.
The number operator is defined as $\hn_r=\had_r\ha_r$.
We denote by $\vac$ the state without any bosons, i.e., the unique normalized state such that $\ha_r\vac=0$ for any $r\in\La$.
For each $x\in\calE$, we define
\eq
\hd_x:=\zeta\ha_x+\sum_{u\in\calN(x)}\ha_u,
\en
where $\zeta>0$ is a model parameter, and $\calN(x)$ is the set of sites in $\calI$ that are on the bonds connected to $x$.

We consider the Hamiltonian $\hH=\Hhop+\Hint$ with the hopping Hamiltonian
\eq
\Hhop:=t\sum_{x\in\calE}\hdd_x\hd_x,
\lb{Hh}
\en
where $t>0$, and the interaction Hamitlonian
\eq
\Hint:=\frac{U}{2}\sum_{u\in\calI}\hn_u(\hn_u-1),
\lb{Hi}
\en
where  $U>0$.
Our model is characterized by the three parameters $\zeta$, $t$, and $U$.
Note that $\Hhop$ is rewritten in the standard form as $\Hhop=\sum_{r,s\in\La}t_{r,s}\had_r\ha_s$, where the amplitude $t_{r,s}$ describes hopping between the nearest and some of the next nearest neighbor sites, and on-site potentials.
Note also that $\Hint$ describes on-site repulsive interaction only on sites in $\calI$.

\begin{figure}
\centerline{\epsfig{file=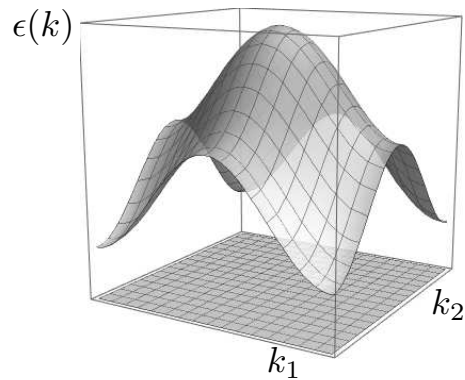,width=6cm}}
\caption[dummy]{
The single particle energy eigenvalue $\epsilon(k)$ as a function of the wave number vector $k=(k_1,k_2)$ for the models defined on the square lattice $(\calE,\calB)$ as in Figure~\ref{f:bd}.
Here $k_j$ ($j=1,2$) runs between $-\pi$ and $\pi$.
There are flat bands (formed by the $\hb$ states) with zero energy  and a dispersive band (formed by the $\hd$ states) with width $4pt$.
The band gap is equal to $\zeta^2 t$.
See Appendix~\ref{a:bands}, in particular, \rlb{dispersion}.
}
\label{f:bands}
\end{figure}

The hopping Hamiltonian $\Hhop$ describes a tight-binding model with a flat band.
To see this we define
\eq
\hb_u:=\frac{1}{\sqrt{2+\zeta^2}}(\zeta\ha_u-\ha_x-\ha_y),
\en
for $u\in\calI$ where $x,y\in\calE$ are the sites connected by the bond corresponding to $u$.
One can easily verify the orthogonality $[\hd_x,\hbd_u]=0$ for any $x\in\calE$ and $u\in\calI$.
This means that $\Hhop\hbd_u\vac=0$ for any $u\in\calI$.
Since $\Hhop\ge0$ and the states $\hbd_u\vac$ with $u\in\calI$  are linearly independent, we see that $\Hhop$ has $|\calI|$ independent single-particle ground states (with zero energy).
See Figure~\ref{f:bands} and Appendix~\ref{a:bands} for more about the single-particle energy eigenvalues.

Let us state our first theorem, which characterizes the ground state of the model.

\prop{Theorem 1}
Consider the above model with particle number $N=|\calI|$.
For any $\zeta>0$, $t>0$, and $U>0$, the ground state of $\hH$ is unique, has vanishing energy, and is written as 
\eq
\GS=(\prod_{u\in\calI}\hbd_u)\vac.
\lb{GS}
\en

Note that, in the ground state \rlb{GS}, exactly one particle is associated with adjacent three sites including $u\in\calI$.
This suggests that the particles are distributed almost uniformly over the lattice, and the state resembles the ground state of a ``solid'' or, more precisely, a Mott insulator.
But the property of the ground state turns out to be much richer than a simple Mott insulator.
In fact we shall argue in section~\ref{s:OPN} that the ground state does not correspond to a genuine Mott insulator when it exhibits (quasi) ODLRO.

The proof of Theorem 1 is an easy application of the technique developed in \cite{Tasaki92e,MielkeTasaki1993,Tasaki1998B} for the (fermionic) Hubbard model.

\prop{Proof of Theorem 1}
It follows from the definitions that $\hH\GS=0$.  Since $\hH\ge0$, this proves that $\GS$ is a ground state.
We only need to prove that it is the unique ground state.

We first note that any single-particle state can be expressed by a linear combination of the operators $\hbd_u$ with $u\in\calI$ and $\hdd_x$ with $x\in\calE$.
This follows by observing that $\hbd_u$ and $\hdd_x$ are all linearly independent, and there are exactly $|\calI|+|\calE|=|\La|$ operators.
We thus see that any $N$ particle state is written as a linear combination of the basis states $\Phin=\{\prod_{u\in\calI}(\hbd_u)^{n_u}\}\{\prod_{x\in\calE}(\hdd_x)^{n_x}\}\vac$, where the ``occupation number'' $\bsn=(n_r)_{r\in\La}$ satisfies $\sum_{r\in\La}n_r=N$.

Our proof is based on the standard argument for frustration free Hamiltonians.
Let $\ket{\Psi}$ be an arbitrary state with $N=|\calI|$ particles such that $\hH\ket{\Psi}=0$, and expand it as $\ket{\Psi}=\sum_{\bsn}\alpha_{\bsn}\Phin$.
Noting that $\hdd_x\hd_x\ge0$ and $\hn_u(\hn_u-1)=(\had_u)^2(\ha_u)^2\ge0$, we see that  $\hdd_x\hd_x\ket{\Psi}=0$ and $(\had_u)^2(\ha_u)^2\ket{\Psi}=0$  for all $x\in\calE$ and $u\in\calI$.
These relations further imply that $\hd_x\ket{\Psi}=0$ for all $x\in\calE$ and $(\ha_u)^2\ket{\Psi}=0$ for all $u\in\calI$.
The first condition implies that $\alpha_{\bsn}=0$ whenever $n_x\ne0$ for some $x\in\calE$.
Thus the ground state contains only the $\hat{b}$-states.
Then the second condition implies that  $\alpha_{\bsn}=0$ whenever $n_u\ge2$ for some $u\in\calI$.
This means that $\ket{\Psi}$ is a constant multiple of  $\GS$.~\qedm

\section{Ground state phase transition}%
We first derive an exact loop-gas model representation of the ground state.
Further details of the derivation are given in Appendix~\ref{A:LGR}.
Note that in this quantum-classical correspondence, unlike in the standard correspondence via path-integral, a $d$-dimensional quantum state is mapped to a $d$-dimensional classical system.
This is a peculiar point about our model, and is analogous to the loop-gas representations of the Affleck-Kennedy-Lieb-Tasaki (AKLT) model \cite{KLT} and the Kitaev model \cite{LeeKOK2019}.

\begin{figure}
\centerline{\epsfig{file=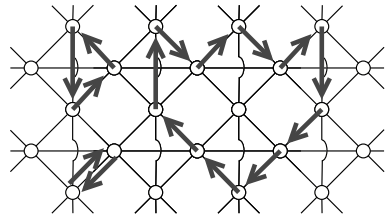,width=5cm}}
\caption[dummy]{
A typical configuration of loops on $\calI$ corresponding to the lattice in Fig.~\ref{f:bd}~(a).
By an arrow from site $u$ to $v$ we indicate the commutator $[\hb_u,\hbd_v]$.}
\label{f:loops}
\end{figure}

Note first that the $\hb$-operators satisfy the commutation relations
\eq
[\hb_u,\hbd_v]=\begin{cases}
1&u=v\\2\beta&u\approx v\\\beta&u\sim v\\0&\text{otherwise},
\end{cases}
\lb{bb}
\en
where 
\eq
\beta:=(2+\zeta^2)^{-1}.
\en
Here $u\approx v$ indicates that the bonds corresponding to $u$ and $v$ connect the same pair of sites, and  $u\sim v$ indicates that the bonds for $u$ and $v$ share a single common site.
By repeatedly using \rlb{bb}, one finds that the normalization factor $\langle\Phi_{\rm GS}\GS$ is represented as \cite{Kimchi}
\eq
\langle\Phi_{\rm GS}\GS=\sum_{\calL}\beta^{|\calL|},
\lb{GG}
\en
where the sum is over all possible sets $\calL=\{\ell_1,\ldots,\ell_n\}$ with $n=0,1,2,\ldots$ of oriented loops.
See Fig.~\ref{f:loops}.
By an oriented loop of length $m$, we mean a sequence $\ell=(u_1,\ldots,u_m)$ of $m$ distinct sites in $\calI$ such that $u_j\approx u_{j+1}$ or $u_j\sim u_{j+1}$ for $j=1,\ldots,m$, where we set $u_{m+1}=u_1$.
Here we identify the new sequence obtained by the shift $u_j\to u_{j+1}$ for $j=1,\ldots,m$ with the original sequence.
This means that, for $u_1,u_2\in\calI$ such that $u_1\sim u_2$, there is a unique loop, $\ell=(u_1,u_2)$, of length two, while for  $u_1,u_2,u_3\in\calI$ such that $u_1\sim u_2$, $u_2\sim u_3$, and $u_3\sim u_1$, there are two loops, $\ell=(u_1,u_2,u_3)$ and $\bar{\ell}=(u_3,u_2,u_1)$, of length three.
We also assume that, in any set $\calL=\{\ell_1,\ldots,\ell_n\}$, no loops share a common site.
To take into account the factor 2 in \rlb{bb}, we interpret $u,v\in\calI$ with $u\approx v$ as being  connected via two distinct paths, and properly over-count loops according to this interpretation.
Finally we wrote $|\calL|=\sum_{j=1}^n|\ell_j|$, where $|\ell|$ denotes the length of a loop $\ell$.
See Appendix~\ref{A:LGR} for details.

We can derive a similar representation for correlation functions.
Noting that $\ha_u\GS=\zeta\sqrt{\beta}(\prod_{w\ne u}\hbd_w)\vac$ for $u\in\calI$, we find that
\eq
\GSb\had_v\ha_u\GS=\zeta^2\beta\sideset{}{'}\sum_{\calL,\,\omega:u\to v}\beta^{|\calL|+|\omega|},
\lb{GaaG}
\en
where $\calL$ is again summed over sets of oriented loops, and  $\omega$ is summed over all self-avoiding walks connecting $u$ to $v$, i.e., a sequence $\omega=(u_0,\ldots,u_m)$ of $m+1$ distinct sites in $\calI$ such that $u_0=u$, $u_m=v$, and $u_j\approx u_{j+1}$ or $u_j\sim u_{j+1}$ for $j=0,\ldots,m-1$.
The length of the walk is defined as $|\omega|=m$.
The prime in the sum indicates that $\omega$ and loops in $\calL$ do not share common sites.
By combining \rlb{GG} and \rlb{GaaG}, we see that the off-diagonal correlation is represented as
\eq
\bktgs{\had_v\ha_u}=\frac{\GSb\had_v\ha_u\GS}{\langle\Phi_{\rm GS}\GS}
=\frac{\zeta^2\beta\sum'_{\calL,\,\omega:u\to v}\beta^{|\calL|+|\omega|}}{\sum_{\calL}\beta^{|\calL|}}.
\lb{aa}
\en
The correlation function $\bktgs{\had_r\ha_s}$ with general $r,s\in\Lambda$ has a similar (but slightly more complicated) representation, and should behave almost similarly as $\bktgs{\had_v\ha_u}$, especially when the distance between $r$ and $s$ is large.

The above summations over loops and walks are in general nontrivial and hardly evaluated explicitly.
When the basic lattice $(\calE,\calB)$ is a chain with  bonds connecting nearest neighbor sites, i.e., $d=p=1$ in the notation of section~\ref{s:def}, one can evaluate the summations by using, e.g., the transfer matrix method to show for a long enough chain that
\eq
\bktgs{\had_v\ha_u}=C\Bigl(\frac{2\beta}{1+\sqrt{1+4\beta^2}}\Bigr)^{|u-v|}\le C\,2^{-|u-v|},
\en
where $C=\zeta^2 \beta/\sqrt{1+4\beta^2}$.
The correlation always decays exponentially, and the ground state is disordered.

Let us turn to the models in higher dimensions.
When $\beta$ is small (or $\zeta$ is large), we can easily prove that  the corresponding loop-gas model is in the disordered phase, where configurations with only small loops are dominant.
This corresponds to a disordered ground state that describes a Mott insulator.
To see this, we relax the constraint $\calL\cap\omega=\emptyset$ in \rlb{GaaG} to get
\eqa
\GSb\had_v\ha_u\GS&\le\zeta^2\beta\sum_{\calL,\,\omega:u\to v}\beta^{|\calL|+|\omega|}
\nl&=\zeta^2\beta\langle\Phi_{\rm GS}\GS\sum_{\omega:u\to v}\beta^{|\omega|}.
\ena
This implies 
\eq
\bktgs{\had_v\ha_u}\le\zeta^2\beta\sum_{\omega:u\to v}\beta^{|\omega|}
=\zeta^2\beta\hspace{-3mm}\sum_{n=\operatorname{dist}(u,v)}^\infty\hspace{-3mm}\Omega_{u,v}(n)\,\beta^n,
\lb{aabn}
\en
where $\Omega_{u,v}(n)$ is the total number of self-avoiding walks of length $n$ that connect $u$ and $v$, and $\operatorname{dist}(u,v)$ is the minimum length of such walks.
For $u\in\calI$, let $s_u$ and $d_u$ be the numbers of $v$ such that $u\sim v$ and $u\approx v$, respectively.
We let $\nu$ be a constant such that $s_u+2d_u\le\nu+1$ for any $u\in\calI$.
Then we have $\Omega_{u,v}(n)\le(\nu+1)\nu^{n-2}$.
({\em Proof:}\/ There are  at most $(\nu+1)$ choices for the first step, and at most  $\nu$ choices for each of the following $n-2$ steps.
There is no choice in the final step because the walk ends at $v$.)
This bound, with \rlb{aabn}, implies that the correlation decays exponentially for sufficiently small $\beta$, or, equivalently, sufficiently large $\zeta$.

\prop{Theorem 2}
Let $\zeta$ be such that $\nu\beta=\nu/(2+\zeta^2)<1$.
Then we have for any $u,v\in\calI$ that
\eq
\bktgs{\had_v\ha_u}\le \frac{\zeta^2\beta(\nu+1)}{\nu^2\,(1-\nu\beta)}(\nu\beta)^{\operatorname{dist}(u,v)}.
\lb{aaUB}
\en

When  the dimension is larger than one and $\beta$ is sufficiently large, on the other hand, it is expected that the loop-gas model is in the percolating phase where a macroscopically large loop (or a walk) appears.
The transition can be characterized by the divergence of the ``static structure factor''
\eq
S := (\zeta^2\beta)^{-1}\sum_{v\in\calI}\bktgs{\had_v\ha_u}.
\lb{StructureFactor}
\en
If the critical value of $\beta$ for the transition is less than $1/2$, which is the upper bound for $\beta$, the ground state undergoes a phase transition.

Let us first examine the standard large-$d$ (or mean-field) approximation, in which one fixes $\tbeta=\nu\beta$ and lets $d\up\infty$ (and hence $\nu\up\infty$).
It is well-known that, for $\tbeta<1$, one can neglect contributions from loops in this limit.
See Appendix~\ref{A:ID}.
Then $S$ is evaluated for any $\tbeta<1$  as
\eq
S\simeq\sum_{\omega:u\to\cdot}\beta^{|\omega|}=\sum_{n=0}^\infty\nu^n\beta^n=\frac{1}{1-\tbeta},
\lb{DIW}
\en
where $\omega$ is summed over all self-avoiding walks (with an arbitrary length) that starts from $u$.
We here noted that the number of walks is almost equal to $\nu^n$.
The structure factor $S$ diverges as $\beta$ approaches the critical value $\nu^{-1}$.

The existence of a phase transition is also suggested by a more careful examination of the loop-gas model.
Recall that, in \rlb{GG} or \rlb{GaaG}, every loop of length three or more is summed exactly twice with different orientations.
This is equivalent to considering unoriented loops, but with an extra factor 2 for each loop.
We thus see that our loop-gas model resembles that obtained from the high-temperature expansion of an O(2) symmetric classical ferromagnetic spin system at a finite temperature.
The O(2) symmetry corresponds to the U(1) phase symmetry in the original quantum system.
From this analogy we conjecture that in dimensions three or higher the ground state $\GS$ may exhibit BEC  (where $\bktgs{\had_v\ha_u}$ has long-range order)  while in two dimensions it may exhibit a quasi BEC (where  $\bktgs{\had_v\ha_u}$ shows power law decay), both at sufficiently large $\beta$.

To prove the existence of BEC is extremely difficult, if not impossible, since we need to show the breakdown of a continuous symmetry.
See \cite{ALSSY} and references therein for rare cases where the existence of BEC can be established rigorously by using the method of reflection positivity.
(A readable account of the reflection positivity method can be found in  \cite{TasakiBook}.)
In the present model, we have a rigorous result only for the system defined on a tree, where one can apply  standard techniques (see, e.g., \cite{Thompson,AKLT}).
See Appendix~\ref{A:tree}.

\section{Numerical evidence of a Kosterlitz-Thouless transition}
To see whether the ground state exhibits a phase transition,
we carried out a Monte Carlo simulation of the corresponding loop-gas model by using the worm algorithm \cite{Worm}.
We focus on the two-dimensional models constructed from the $L\times L$ square lattice with bond number $p=1$ and 2 and with periodic boundary conditions.  See Fig.~\ref{f:bd}.

A central quantity that we measured is the static structure factor $S$ defined in \rlb{StructureFactor}.
When the correlation $\langle \hat a_v^{\dagger} \hat a_u \rangle_{\rm GS}$
becomes long-ranged, the structure factor should diverge as a function of $L$.
Specifically, $S= \mathcal{O}(L^d)$ if the system possesses ODLRO,
and $S= \mathcal{O}(L^{d-\eta})$ if it has quasi-ODLRO characterized by the
correlation-decay exponent $\eta$, while
$S= \mathcal{O}(1)$ in the disordered phase.
In particular, $\eta = 1/4$ at the KT transition point
with the multiplicative logarithmic correction of $(\log L)^{1/8}$
\cite{Kosterlitz1974}.

To capture more specific features of the KT transition,
we also measured the helicity modulus,
which is proportional to the super-fluid density.
It can be computed as the Monte Carlo average of
the squared total winding number of all loops \cite{Pollock1987,Kimchi},
\eq
\Upsilon :=
\biggl\langle \Bigl( \sum_{\ell \in \calL} w_\ell \Bigr)^2 \biggr\rangle_{\rm\!\!MC} = 
\biggl\langle \sum_{\ell \in \calL} w_\ell^2 \biggr\rangle_{\rm\!\!MC},
\en
where the summation is over all loops in the loop-gas configuration $\calL$
and $w_l$ is the winding number of the loop $l$ around the horizontal direction
of the lattice.
As we pass the KT transition point entering the quasi-ODLRO phase,
this quantity is expected to show a discontinuous jump
from zero to the universal value $2/\pi$,
and keep increasing afterwards.

\begin{figure}[b]
  \centerline{\epsfig{file=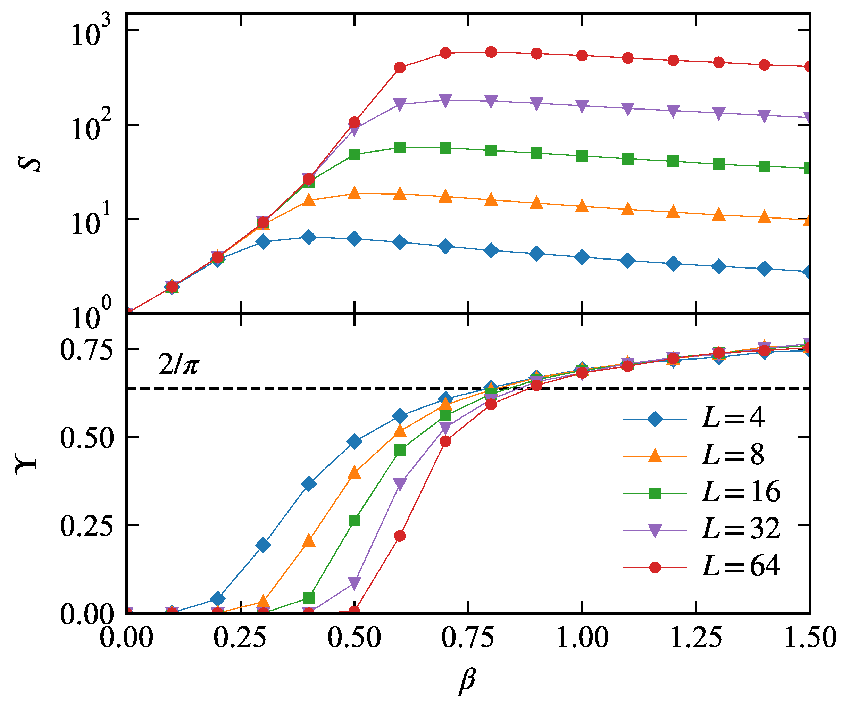,width=8cm}}
  \caption[dummy]{
    The static structure factor $S$ (top) and the helicity modulus (bottom)
    as functions of $\beta$ for $p=1$.
    The horizontal line in the bottom panel indicates
    the universal jump $2/\pi$, the thermodynamic value
    expected for the KT transition point.
  }
  \label{f:BetaDependenceP1}
\end{figure}

Let us start from the simplest model with $p=1$.
The top panel of Fig.~\ref{f:BetaDependenceP1} shows the $\beta$ dependence of
the static structure factor $S$
for varying system sizes up to $L=64$.
Clearly we do not observe
any singularity in the region $\beta \le 1/2$, where the classical-quantum correspondence is valid.
The data, however, seem to indicate that the system, as a classical
loop-gas model, has a phase transition beyond $\beta=1/2$.

The bottom panel of Fig.~\ref{f:BetaDependenceP1} shows the $\beta$ dependence of
the helicity modulus $\Upsilon$.
It can be seen from the figure that $\Upsilon$ goes from zero to some finite
value greater than the expected jump, indicating that a KT transition takes place near $\beta \sim 1$.
This is in contrast to the case studied in \cite{Kimchi}.
We shall recall, however, that the loop-gas model with $\beta \sim 1$ does not correspond to a quantum mechanical ground state.

\begin{figure}
\centerline{\epsfig{file=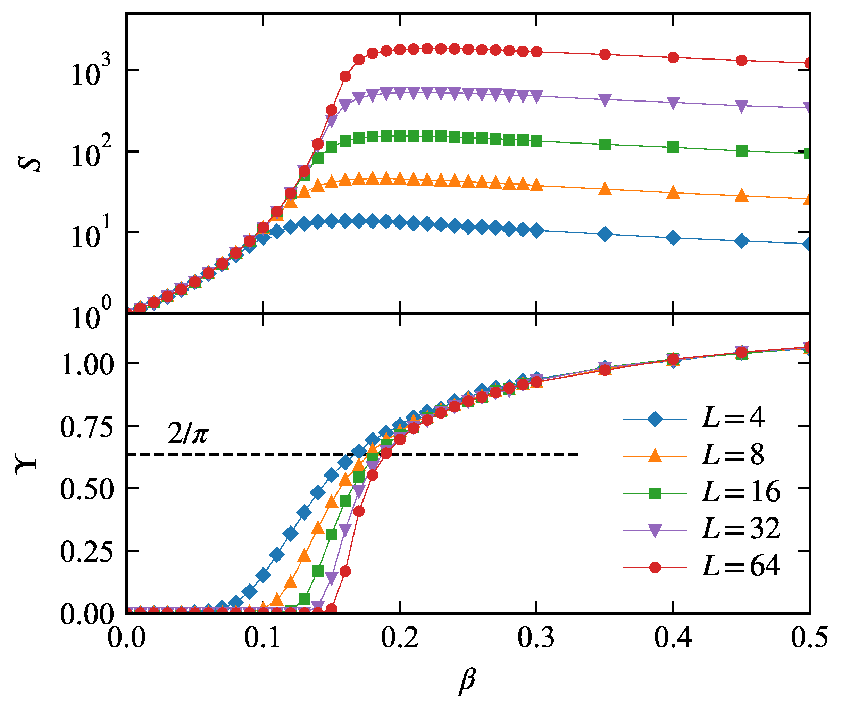,width=8cm}}
\caption[dummy]{
  The static structure factor $S$ (top) and the helicity modulus (bottom)
  as functions of $\beta$ for $p=2$.
  The horizontal line in the bottom panel indicates
  the universal jump $2/\pi$, the thermodynamic value
  expected for the KT transition point.
}
\label{f:BetaDependence}
\end{figure}

Having observed that the model with $p=1$ does not exhibit a transition (in the range $\beta<1/2$ that is relevant for us), let us focus on the model with $p=2$.
Figure \ref{f:BetaDependence} shows the $\beta$ dependence of
the structure factor (top) and the helicity modulus (bottom) for varying system sizes up to $L=64$ in the loop-gas model corresponding to $p=2$.
Clearly the  static structure factor $S$ shows a diverging behavior for large but not too large values of $\beta$.
The helicity modulus $\Upsilon$ also shows an increase from zero to some finite
value greater than the expected universal jump $\pi/2$, strongly suggesting the presence of a KT transition in the physically meaningful region $\beta<1/2$.

\begin{figure}
\centerline{\epsfig{file=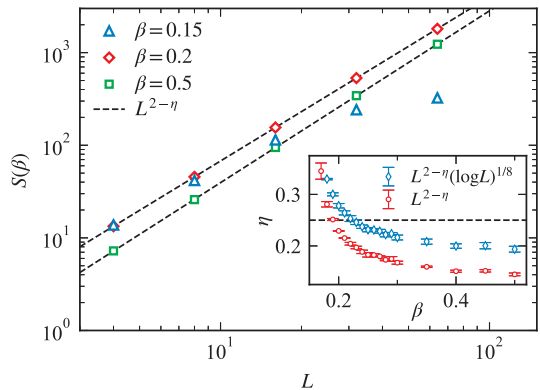,width=7.5cm}}
\caption[dummy]{
  The size dependence of the structure factor $S$.
  The dashed lines are the power function fitting.
  The inset shows the estimates of the critical exponent $\eta$ obtained
  by fitting with the simple power function (blue) and
  the power function with the expected logarithmic correction (red).
}
\label{f:Eta}
\end{figure}

Further evidence of a transition can be obtained from the size-dependence of
the structure factor, shown in Fig.~\ref{f:Eta}.
For small values of $\beta$, before the slope of the curve reaches 7/4,
it is not a straight line but bends and starts converging
to a finite value, as seen for $\beta = 0.15$.
As we increase $\beta$, the curve becomes a straight line
with slope approximately equal to $7/4$ ($\beta = 0.20$ in the figure).
Further increasing of $\beta$ brings the slope of the straight line
greater than $7/4$ ($\beta=0.5$ in the figure).
To estimate the slope of the straight part of the curve more quantitatively,
we analyzed the data with or without the expected logarithmic correction,
i.e., fit the form $L^{2-\eta}$ or $(\log L)^{1/8} L^{2-\eta}$ to the data
regarding the exponent $\eta$ as a fitting parameter,
although according to \cite{Janke1997}
it is not clear whether including such a logarithmic correction
would improve the estimates or not.
The inset shows the result of the fitting with or without the logarithmic factor.
From this result, we may conclude that the KT transition takes place
at $\beta = 0.20(1)$,
which is consistent with the universal jump in the bottom panel
of Fig.~\ref{f:BetaDependence} considering the expected slow ($\sim (\log L)^{-1}$)
convergence of $\Upsilon$ at the transition point.

\begin{figure}
\centerline{\epsfig{file=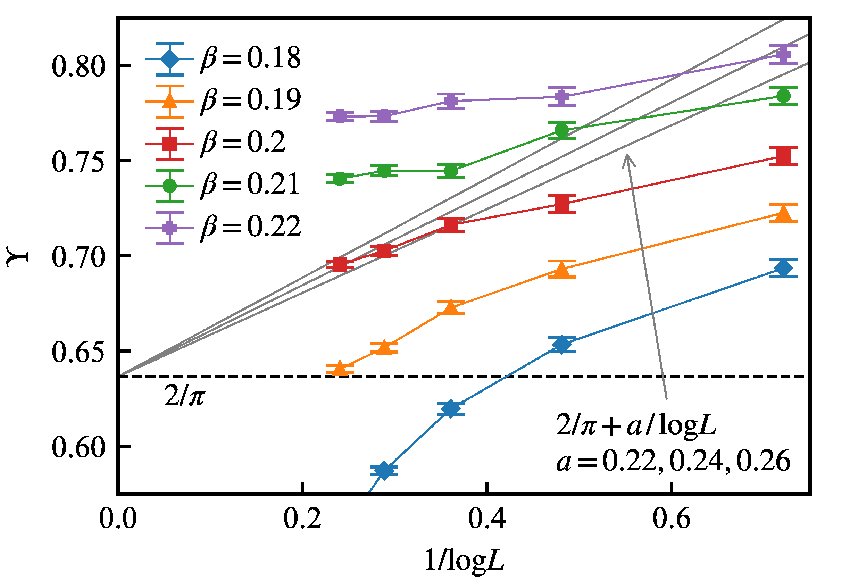,width=8cm}}
\caption[dummy]{
  The size-dependence of the helicity modulus near the
  transition point for $p=2$. Straight lines are guide for the eye.
}
\label{f:HelicityModulus_vs_Size}
\end{figure}

To get an additional evidence of a KT transition,
we examined this size-dependence of the helicity modulus
near the transition point in more detail.
As shown in Fig.\ref{f:HelicityModulus_vs_Size},
we plot $\Upsilon$ as a function of $1/\log L$ around the
critical value of $\beta$.
At $\beta = 0.2$, the value close to the critical point
estimated from the size dependence of the structure factor,
the data is consistent with the linear convergence in $1/\log L$
to the universal jump, as predicted in \cite{WeberMinnhagen}
as a characteristic size-dependence for the KT trnasition.
This may be taken as another evidence suggesting the KT nature
of the transition.

\section{The model with other particle numbers}
\label{s:OPN}
Let us briefly discuss the properties of the models with different particle numbers, and related issue about the charge gap.

When the number of particles, $N$, is less than $|\calI|$, we see that the space of the ground states is spanned by $\ket{\Phi_S}=(\prod_{u\in S}\hbd_u)\vac$ where $S$ is an arbitrary subset of $\calI$ such that $|S|=N$.
The ground states show macroscopic degeneracy, which should be immediately lifted when a generic infinitesimal perturbation is added to the Hamiltonian.

The case with $N>|\calI|$ is difficult, and we have almost no exact results.
We can nevertheless construct a class of exact energy eigenstates, which may be regarded as new examples of quantum many-body scars \cite{scar1,scar2}.
Assume that the lattice $(\calE,\calB)$ is connected and bipartite.
To be precise we say that $(\calE,\calB)$ is bipartite if there is a decomposition $\calE=\calE_+\cup\calE_-$ such that two sites $x$, $y$ may be connected by a bond in $\calB$ only when $x\in\calE_+$, $y\in\calE_-$ or $x\in\calE_-$, $y\in\calE_+$.
Let us define 
\eq
\hD:=\sum_{x\in\calE_+}\ha_x-\sum_{x\in\calE_-}\ha_x.
\en
Then, for any subset $S\subset\calI$ and $n=1,2\ldots$, it is easily shown that the state 
\eq
\ket{\Phi_{S,n}}=(\hDD)^n(\prod_{u\in S}\hbd_u)\vac,
\en
which has $|S|+n$ particles, is an energy eigenstate with eigenvalue $E_n=nt\zeta^2$.
See Appendix~\ref{A:ES}.

Let us denote by $\EGS_N$ the ground state energy of the model with $N$ particles, and define the charge gap (or the jump in chemical potential) as 
\eq
\Di_N=\EGS_{N+1}+\EGS_{N-1}-2\EGS_N.
\en
It is believed that the charge gap provides a simple criterion for conductivity in the sense that the ground state is insulating if $\Di_N$ is positive and of order 1 \cite{LiebWu}.

Since we have $\EGS_N=0$ for $N\le|\calI|$ in the present model, we see that $\EGS_{|\calI|+1}$ is nothing but the charge gap $\Di_{|\calI|}$, which is directly relevant to the property of the model with $|\calI|$ particles.
We conjecture that, for $\beta<\betac$, the charge gap $\Di_{|\calI|}$ is strictly positive and our ground state describes a Mott insulator, while, for $\beta>\betac$ where the ground state exhibits (quasi) BEC, $\Di_{|\calI|}$ vanishes in the infinite volume limit according to the theorem in \cite{TW}.
Therefore the ground state, although Mott insulator-like, is not a genuine Mott insulator.

\section{Discussion}
\label{s:discussion}
We proposed a new class of exactly solvable models of interacting bosons with a flat band, and argued that the Mott insulator-like ground states may exhibit (quasi) BEC.
The conjecture is supported by the strong numerical evidence that the two-dimensional model exhibits a KT transition.
The properties of the three-dimensional models remain to be investigated.

We believe it important that a new exactly solvable model that exhibits (or, that is conjectured to exhibit) nontrivial condensation phenomena has been discovered.
It is also interesting to investigate the possibility of similar models of electrons, which should exhibit superconductivity.

It may be counterintuitive that our exact ground state \rlb{GS}, which consists of bosons almost localized at each $u\in\calI$, exhibits ODLRO.
One should note however that the operator $\hbd_u$ creates a coherent superposition of the three states in which a particle is at $x$, $y$, and $u$.
The coherence ``propagates'' in the system thus generating off-diagonal correlation, which may be short-ranged or long-ranged \cite{Kimchi}.
At least mathematically, the situation is parallel to that for the long-range N\'eel order in the exact valence-bond ground states of the AKLT model in high dimensions \cite{KLT,AAH,AKLT}.

Indeed this point is related to the fact that the states proposed in \cite{Kimchi} only have short-ranged off-diagonal correlation while our ground state on a suitable two-dimensional lattice exhibits quasi  ODLRO.
The basic difference is not of qualitative but of quantitative nature, namely, the commutator $[\hb_u,\hbd_v]$ of neighboring sites, which give the basic parameter $\beta$, and the coordination number $\nu+1$ can be larger in our models compared with that in \cite{Kimchi}.

It is worth recalling that a two-dimensional system of interacting bosons at zero temperature generically exhibits genuine ODLRO rather than quasi ODLRO. 
We believe that the present model exhibits only quasi ODLRO because of its peculiar ground state structure \rlb{GS} expressed only in terms of local bosonic operators $\hbd_u$.
This is consistent with the fact that the ground state of the present two-dimensional model is represented in terms of a classical statistical mechanical model (i.e., the loop-gas model) in two dimensions, while a ground state in two-dimensional quantum system generally corresponds to a three-dimensional classical system.
We conjecture that the observed quasi ODLRO will be immediately elevated into genuine ODLRO when the model is perturbed.
See, e.g., \cite{Nomura} for the discussion of a similar behavior in the ground state correlation function of the one-dimensional AKLT model.

\begin{figure}[b]
\centerline{\epsfig{file=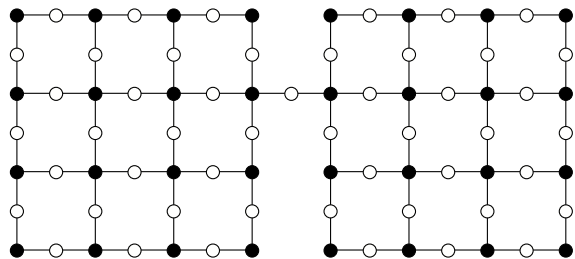,width=7.5cm}}
\caption[dummy]{
The lattice obtained by connecting two lattices based on the square lattice  by a single bond.
The number of particles in each sublattice is almost constant in the corresponding ground state. 
}
\label{f:coupled}
\end{figure}

We have repeatedly stressed that our exact ground state \rlb{GS} has a Mott-insulator like character since it is generated by local operators $\hbd_u$.
Its peculiar property is most clearly seen if one considers a special lattice $(\calE,\calB)$ obtained by connecting two arbitrary standard lattices by a single bond.
See Fig.~\ref{f:coupled}.
In the ground state of the model defined on the corresponding decorated lattice $\La$, the number of particles in each of the two sublattices is almost constant.  (To be precise, it fluctuates only by one.)
Note that this is also true when we properly choose the lattice so that the ground state exhibits (quasi) ODLRO.
Such a ground state with (quasi) ODLRO and vanishing particle number fluctuation is quite exotic since (quasi) ODLRO is usually accompanied by large density fluctuation.
We should not, however, jump to the conclusion that a novel exotic phase of matter has been discovered.
It is possible that the zero fluctuation is a singular property of the exactly solvable model, and normal large fluctuation is recovered once the model is perturbed.
For the moment we believe that this (less exciting) scenario is plausible.
We nevertheless stress that the discovery of models in which anomalously small particle number fluctuation and (quasi) ODLRO may coexist indicates that strongly interacting systems of bosons may exhibit unexpectedly rich behavior.

\begin{acknowledgments}
It is a pleasure to thank  David Huse, Kensuke Tamura, and Itamar Kimchi for valuable comments.
H.K. was supported by JSPS Grant-in-Aid for Scientific Research on Innovative Areas No. JP20H04630, JSPS Grant-in-Aid for Scientific Research No.~JP18K03445, and the Inamori Foundation.
N.K. was supported by JSPS Grant-in-Aid for Scientific Research No.~JP19H01809.
The numerical calculation is done on System B (ohtaka) of ISSP Supercomputer Center, The University of Tokyo.
\end{acknowledgments}

\begin{widetext}

\appendix
\section{Single-particle energy eigenvalues}
\label{a:bands}
Let us examine the band structure determined by the hopping Hamiltonian \rlb{Hh}.
We first recall that the states $\hbd_u\vac$ with $u\in\calI$ and $\hdd_x\vac$ with $x\in\calE$ form a basis of the single-particle Hilbert space.
Since the state $\hbd_u\vac$ has zero energy, we focus on states spanned by $\hdd_x\vac$.

Here, for simplicity, we assume that any pair $x,y\in\calE$ is either connected by $p$ distinct bonds in $\calB$ or not connected at all.
We write $x\sim y$ when $x$ and $y$ are connected by $p$ bonds, and denote by $\calE(x)$ the set of $y\in\calE$ such that $x\sim y$.
Recalling the definition
\eq
\hd_x=\zeta\ha_x+\sum_{u\in\calN(x)}\ha_u,
\en
we see that
\eq
[\hd_x,\hdd_y]=\begin{cases}
\zeta^2+pz_x&x=y;\\
p&x\sim y;\\
0&\text{otherwise},
\end{cases}
\lb{dd}
\en
where $z_x=|\calE(x)|$ is the coordination number (i.e., the number of neighboring sites) of the original lattice $(\calE,\calB)$.
Note that $|\calN(x)|=pz_x$.
The commutation relation, along with the definition \rlb{Hh} of $\Hhop$, implies
\eq
[\Hhop,\hdd_x]=(\zeta^2+pz_x)t\,\hdd_x+pt\sum_{y\in\calE(x)}\hdd_y.
\lb{Hhd}
\en

Consider and arbitrary state spanned by $\hd$ operators  
\eq
\ket{\varphi}=\sum_{x\in\calE}\varphi_x\hdd_x\vac,
\en
where $\varphi_x\in\bbC$.
By using the commutation relation \rlb{Hhd} and $\Hhop\vac=0$, we see that
\eqa
\Hhop\ket{\varphi}&=\sum_{x\in\calE}\varphi_x\Bigl\{(\zeta^2+pz_x)t\,\hdd_x+pt\sum_{y\in\calE(x)}\hdd_y\Bigr\}\vac
\nl&=\sum_{x\in\calE}\Bigl\{(\zeta^2+pz_x)t\,\varphi_x+pt\sum_{y\in\calE(x)}\varphi_y\Bigr\}\hdd_x\vac,
\ena
where we switched the roles of $x$ and $y$ to get the final expression.
Thus the Schr\"odinger equation 
\eq
\Hhop\ket{\varphi}=\epsilon\ket{\varphi}\en
reduces to 
\eq
(\zeta^2+pz_x)t\,\varphi_x+pt\sum_{y\in\calE(x)}\varphi_y=\epsilon\,\varphi_x\quad\text{for any $x\in\calE$},
\lb{SheqS}
\en
which is nothing but the standard tight-binding Schr\"odinger equation with hopping $pt$ and on-site potential $(\zeta^2+pz_x)t$.

Now we suppose that $(\calE,\calB)$ is the $d$-dimensional $L\times\cdots\times L$ hypercubic lattice with periodic boundary conditions, i.e.,
\eq
\calE=\{(x_1,\ldots,x_d)\,|\,x_j=1,\ldots,L\},
\en
and $x\sim y$ when $|x-y|=1$.
The coordination number is $z_x=2d$ for all $x$.
Let us take the standard plane wave $\varphi^{(k)}_x=e^{ik\cdot x}$ with $k\cdot x=\sum_{j=1}^dk_jx_j$ and $k\in\calK$, where  the set of wave number vectors is
\eq
\calK=\{(k_1,\ldots,k_d)\,|\,k_j=\frac{2\pi}{L}n_j,\  n_j= -L/2+1,\ldots,L/2\}.
\en
Substituting $\varphi^{(k)}_x$ into \rlb{SheqS}, one readily confirms that this is an energy eigenstate with eigenvalue
\eq
\epsilon(k)=\zeta^2t+2pt\sum_{j=1}^d(\cos k_j+1).
\lb{dispersion}
\en
See Figure~\ref{f:bands} for the case with $d=2$.

\section{Loop-gas representations}
\label{A:LGR}
Let us describe in some detail the derivations of the loop-gas representations \rlb{GG} and \rlb{GaaG}.
From $\hb_u\vac=0$, we have the standard relation
\eq
\langle\Phi_{\rm GS}\GS=\vacb\bigl(\prod_{u\in\calI}\hb_u\bigr)\bigl(\prod_{u\in\calI}\hbd_u\bigr)\vac
=\sum_\per\prod_{u\in\calI}[\hb_u,\hbd_{\per(u)}],
\lb{GG2}
\en
where $\per$ is summed over all permutations of the elements of $\calI$.

\medskip

We first focus on the simper class of models where no sites $u,v\in\calI$ satisfy $u\approx v$.

To see that \rlb{GG2} leads to the claimed representation \rlb{GG}, it is best to first examine a simple example.
Suppose that $\calI=\{u_1,u_2,u_3\}$, and it holds that $u_1\sim u_2$, $u_2\sim u_3$, and $u_3\sim u_1$.
Then we see explicitly from \rlb{GG2} that
\eqa
\langle\Phi_{\rm GS}\GS&=1+[\hb_1,\hbd_2]\,[\hb_2,\hbd_1]+[\hb_2,\hbd_3]\,[\hb_3,\hbd_2]+[\hb_3,\hbd_1]\,[\hb_1,\hbd_3]
+[\hb_1,\hbd_2]\,[\hb_2,\hbd_3]\,[\hb_3,\hbd_1]+[\hb_1,\hbd_3]\,[\hb_3,\hbd_2]\,[\hb_2,\hbd_1]
\nl&=1+3\beta^2+2\beta^3,
\ena
where we abbreviated $\hb_{u_j}$ as $\hb_j$.

To treat general cases, fix $\per$ and let $\calI'=\{u\in\calI\,|\,\per(u)\ne u\}$.
It is well-known (and easily proved) that $\calI'$ is decomposed into a disjoint union as $\calI'=\bigcup_{k=1}^nc_k$, where each $c_k$ is a cycle.
A cycle $c$ is a set of more than one site that can be rearranged into an ordered sequnce $(u_1,\ldots,u_m)$ such  that $\per(u_j)=u_{j+1}$ for $j=1,\ldots,m$, where we wrote $u_{m+1}=u_1$.
Defining the weight for the cycle $c$ as $W(c)=\prod_{j=1}^m[\hb_{u_j},\hbd_{u_{j+1}}]$, we see that the summand in  the right-hand side of \rlb{GG2} factorizes as $\prod_{u\in\calI}[\hb_u,\hbd_{\per(u)}]=\prod_{k=1}^nW(c_k)$.
By recalling \rlb{bb}, we get the desired representation \rlb{GG}.

To derive \rlb{GaaG}, we take arbitrary $v\ne u$, and evaluate 
\eq
\GSb\had_v\ha_u\GS=\zeta^2\vacb\bigl(\prod_{w\ne v}\hb_w\bigr)\bigl(\prod_{w\ne u}\hbd_w\bigr)\vac.
\en
Since there is $\hb_u$ but no $\hbd_u$, we must have $[\hb_u,\hbd_{u_1}]$ with $u\sim u_1$ to have a nonzero contribution.
This implies that we also need $[\hb_{u_1},\hbd_{u_2}]$ with $u_1\sim u_2$, and so on.
This is terminated only when we have $[\hb_{u_{m-1}},\hbd_v]$ with $u_{m-1}\sim v$.
We have obtained the contribution from the random walk.  
The contribution from loops can be derived as in the above.

\begin{figure}[b]
\centerline{\epsfig{file=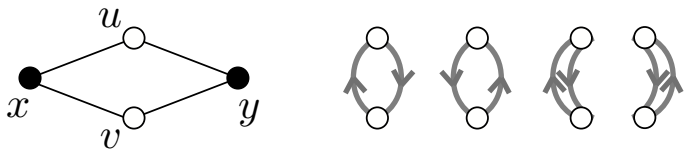,width=9cm}}
\caption[dummy]{
The graphs corresponding to the normalization factor $\langle\Phi_{\rm GS}\GS$ in the simplest model with only two $u,v\in\calI$ such that $u\approx v$.}
\label{f:doublebond}
\end{figure}

\medskip
We now turn to a general class of models where we need to take into account the factor 2 that appears in the commutation relation $[\hb_u,\hbd_v]=2\beta$ for $u,v\in\calI$ such that $u\approx v$.
This can be of course done by redefining the weights for loops and walks, but there is a more elegant way of incorporating the factor.
We use the same definitions for the weights, but regard in general that sites $u$ and $v$ with $u\approx v$ are connected by two distinct paths.

To see that this works, it (almost) suffices to consider the simplest model constructed from the lattice $\calE=\{x,y\}$ with only two sites where $x$ and $y$ are connected by two distinct bonds.
We then let $u$ and $v$ be the sites at the center of these bonds.
We thus have $u\approx v$.
In this case we see from \rlb{GG2} that
\eq
\langle\Phi_{\rm GS}\GS=1+[\hb_u,\hbd_v]\,[\hb_v,\hbd_u]=1+4\beta^2.
\en
The factor 4 is reproduced as in Fig.~\ref{f:doublebond}.

The configurations of loops in these general models can be much more complicated than Fig.~\ref{f:loops} in the main text.
In the model with $d=2$ and $p=2$ depicted in Fig.~\ref{f:bd}~(b), for example, the loops are defined on a lattice similar to Fig.~\ref{f:loops}, but each white site should be replaced by a pair of white sites connected by two distinct paths, and a path connecting two white sites should be replaced by four paths connecting two pairs of white sites.

\section{Large-$d$ approximation}
\label{A:ID}
In the large-$d$ approximation, we can neglect the possibility that a random walk accidentally intersect its trajectory.
This means that the number of length $n$ walks is given by $\nu^n$.
We used this estimate in \rlb{DIW}.
Note that we here do not make distinction between $\nu$ and $\nu+1$.

Let us see why we can neglect the contribution from loops.
Let $\Omega(n)$ be the number of loops that contain a given site in $\calI$.
Since there are at most $(\nu+1)$ choices for the first step, at most $n$ choices for each of the following $n-2$ steps, and no choices in the final step, we have
\eq
\Omega(n)\le(\nu+1)\nu^{n-2}.
\en
Thus the total contribution from all the loops containing a site is bounded from above by
\eq
\sum_{n=2}^\infty\Omega(n)\beta^n\lesssim\sum_{n=2}^\infty\nu^{n-1}\beta^n=\frac{1}{\nu}\sum_{n=2}^\infty\tbeta^n.
\en
This vanishes as $\nu\up\infty$ provided that the sum converges.
({\em Remark:}\/ To be precise, to fix a site and sum over all the loops containing it is not a proper way of evaluating the summations in \rlb{aa}.  But it gives a correct order estimate in terms of $\nu$.)

\section{Bose-Einstein condensation in the model on a tree}
\label{A:tree}

We shall study the model defined on a tree, and show that the ground state exhibits Bose-Einstein condensation for sufficiently small $\zeta$.

Let $(\calE_n,\calB_n)$ be the regular $n$-generation tree with branching number three, as depicted in Fig.~\ref{f:EB}.
As in the main text, we define the corresponding set of sites $\calI_n$, and consider the model of interacting boson on $\calE_n\cup\calI_n$.
Our goal is to show that the ground state exhibits spontaneous symmetry breaking associated with BEC.
We note that the model on the tree with branching number two does not exhibit a phase transition.

\begin{figure}
\centerline{\epsfig{file=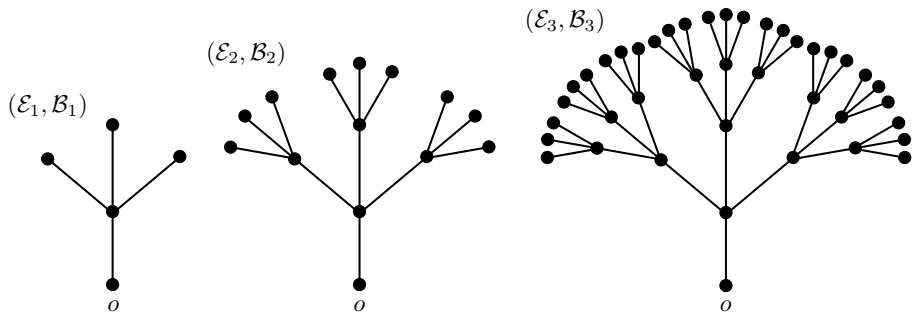,width=12cm}}
\caption[dummy]{
 The first three generations of the tree with branching number three.
 The site at the root is denoted as $o$.}
\label{f:EB}
\end{figure}

A standard method to test for the existence of spontaneous symmetry breaking is to impose boundary conditions that explicitly favor certain order, and see if the effect of the boundary conditions survives in the infinite volume limit.
In the case of ferromagnetic spin systems, this is done by enforcing spins at the boundary to point in a certain fixed direction.
In the case of BEC, the corresponding procedure is to replace $\hbd_u$ at the boundary by $\alpha+\gamma\hbd_u$ with some nonzero $\alpha,\gamma\in\bbC$ and then to examine the expectation value of the annihilation operator deep inside the tree.
In the ferromagnetic Ising model, for example, it is known that the same procedure exactly recovers the result of the Bethe approximation \cite{Thompson}.
See also \cite{AKLT} for a treatment of a quantum spin state.

\begin{figure}
\centerline{\epsfig{file=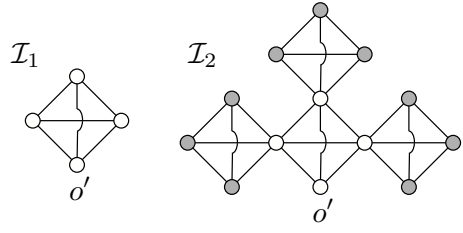,width=6cm}}
\caption[dummy]{
The lattices $\calI_1$ and $\calI_2$.
The bonds denote the connection $u\sim u'$.
The gray circles are the sites in the boundary $\DI_n$, and the root of the lattice is denoted as $o'$.}
\label{f:I}
\end{figure}

To be precise, let $\DI_n$ be the set of sites at the boundary in $\calI_n$.  See Fig.~\ref{f:I}.
We then consider the ground state with plus boundary conditions defined as 
\eq
\GSp=\Bigl(\prod_{u\in\calI_n\backslash\DI_n}\hbd_u\Bigr)\Bigl(\prod_{u\in\DI_n}(1+\hbd_u)\Bigr)\vac.
\en
Note that we have chosen $\alpha=\gamma=1$ for simplicity.
We are interested in the expectation value
\eq
\bktgs{\ha_o}^+=\frac{\GSpb\ha_o\GSp}{\GSpbn\GSp},
\en
especially in its limiting value as $n\up\infty$, where $o$ denotes the site at the root of the tree $(\calE_n,\calB_n)$.

\begin{figure}
\centerline{\epsfig{file=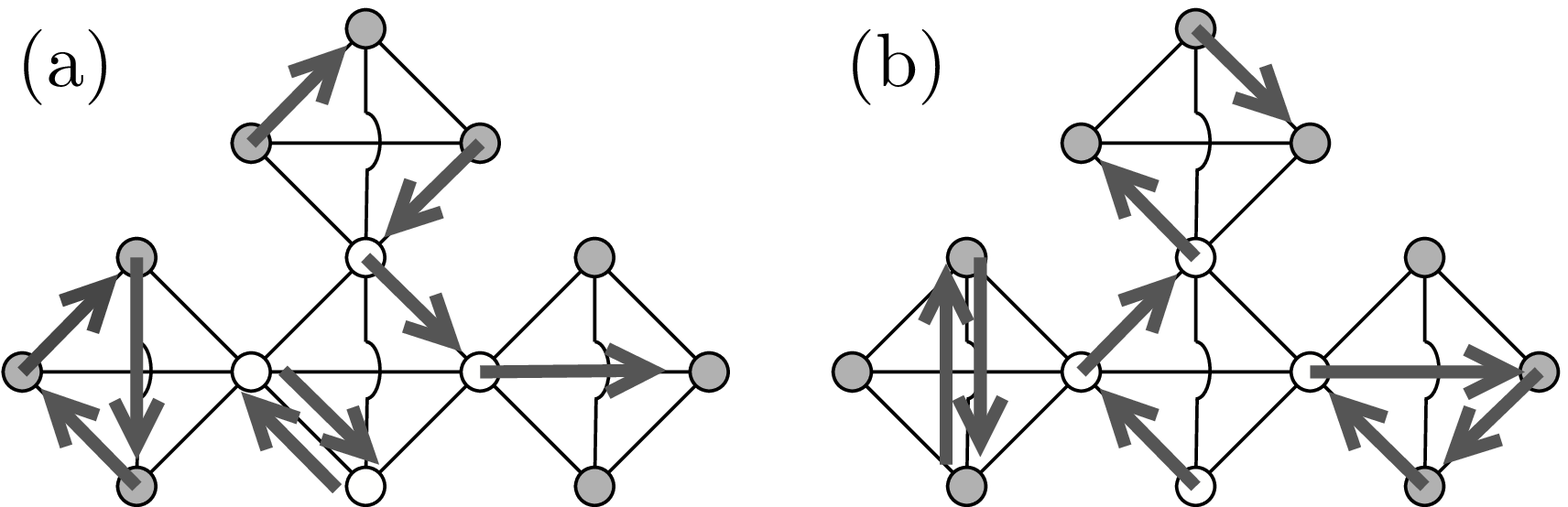,width=9cm}}
\caption[dummy]{Allowed configurations of loops and walks for the representations of (a)~$\GSpbn\GSp$ and (b)~$\GSpb\ha_o\GSp$.
}
\label{f:configs}
\end{figure}

As in the main text, we develop loop-gas representations for $\GSpbn\GSp$ and $\GSpb\ha_o\GSp$.
Reflecting the special geometry of $\calI_n$, the representations contain loops of length two, three, or four, but not larger.
Apart from these loops the representations contain random walks that starts from a site in $\DI_n$ and ends in another site in $\DI_n$.
Note that the symmetry breaking boundary terms play the roles of sources and sinks of the walks.
Of course the loops and the walks should satisfy the site-avoiding conditions.
See Fig.~\ref{f:configs}~(a).
These are all contributions for the representation for $\GSpbn\GSp$, and we have
\eq
\GSpbn\GSp=\sum_{m=0}^\infty\ \sum_{\ell_1,\ldots,\ell_m}\ \sum_{k=0}^\infty\ \sum_{\omega_1,\ldots,\omega_{k}:\DI\to\DI}
\beta^{\sum|\ell_j|+\sum|\omega_{j}|}.
\lb{GGp}
\en

The representation of $\GSpb\ha_o\GSp$ must contain a random walk that starts from the site $o'$,  the root of $\calI_n$, and ends at a site in $\DI_n$.
See Fig.~\ref{f:configs}~(b).
Thus the representation is given by
\eq
\GSpb\ha_o\GSp=\sqrt{\beta}\sum_{\omega_0:o'\to\DI}\ \sum_{m=0}^\infty\ \sum_{\ell_1,\ldots,\ell_m}\ \sum_{k=0}^\infty\ \sum_{\omega_1,\ldots,\omega_{k}:\DI\to\DI}
\beta^{\sum|\ell_j|+\sum|\omega_{j}|}.
\lb{GaGp}
\en

\begin{figure}
\centerline{\epsfig{file=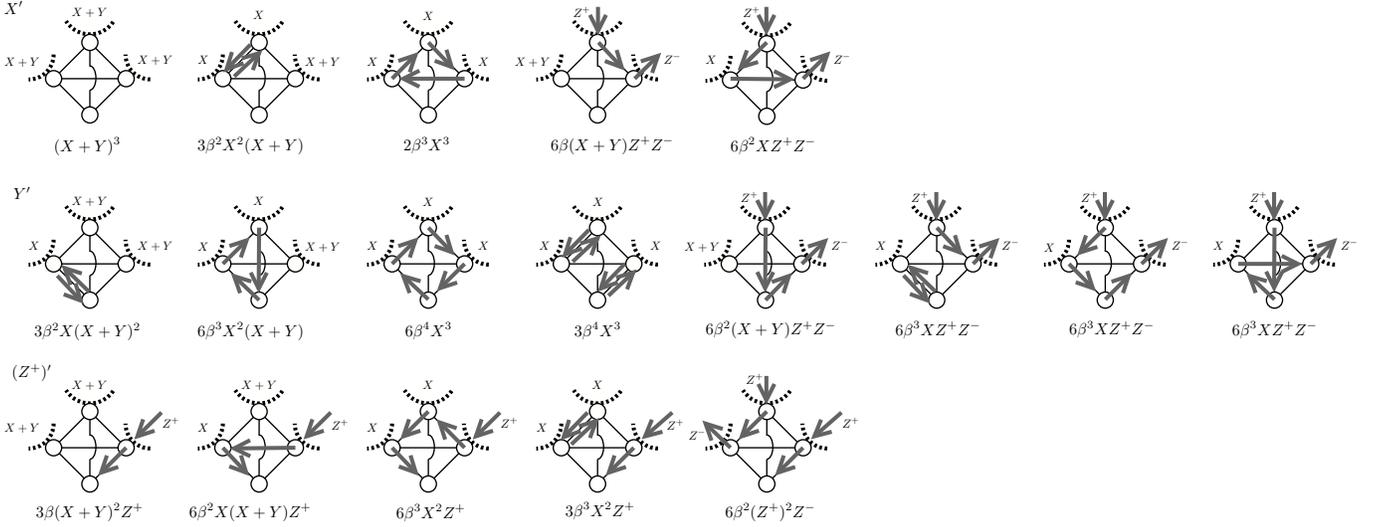,width=18cm}}
\caption[dummy]{
The diagrammatic derivation of the recursion relations  \rlb{r1}, \rlb{r2}, and \rlb{r3}.
$\calI_{n+1}$ is decomposed into the square containing the root $o'$ and three copies of $\calI_n$.
By specifying a configuration (of loops and walks) on the square, possible types of configurations on each branch is determined separately.
By counting the number of similar configurations, we get the recursion relations.
}
\label{f:XYZ}
\end{figure}

Following the standard procedure for models on a tree \cite{Thompson,AKLT}, we shall evaluate these sums by using exact recursion relations.
We define four sums $X_n$, $Y_n$, $Z^+_n$, and $Z^-_n$ of loops and walks as in \rlb{GGp} and \rlb{GaGp} with different condition on the site $o'$ at the root of $\calI_n$.
In $X_n$ we sum over all configurations where no loop or walk touching $o'$.
In $Y_n$ we sum over all configurations where there are two segments (which are part of a loop or a walk) touching $o'$.
In $Z^+_n$ ({\em resp.}\/ $Z^-_n$) we sum over all configurations where there is exactly one segment (which is a  part of a walk) coming into ({\em resp.}\/ going out of) $o'$.
Note that we have $X_0=1$, $Y_0=0$, and $Z^\pm_0=1$.
We see that $\GSpbn\GSp=X_n+Y_n$, $\GSpb\ha_o\GSp=\sqrt{\beta}\,Z^-_n$, and hence
\eq
\bktgs{\ha_o}^+=\sqrt{\beta}\,\frac{Z^-_n}{X_n+Y_n}=\sqrt{\beta}\,\frac{z_n}{1+y_n},
\en
where we defined
\eq
y_n=\frac{Y_n}{X_n},\quad z_n=\frac{Z^-_n}{X_n}.
\en
Now it is straightforward (although tedious) to see that  $X_n$, $Y_n$, $Z^+_n$, and $Z^-_n$ satisfy the exact recursion relations
\eqg
X'=(X+Y)^3+3\beta^2X^2(X+Y)+2\beta^3X^3+6\beta(X+Y)Z^+Z^-+6\beta^2XZ^+Z^-,\lb{r1}\\
Y'=3\beta^2X(X+Y)^2+6\beta^3X^2(X+Y)+6\beta^4X^3+3\beta^4X^3+6\beta^2(X+Y)Z^+Z^-+18\beta^3XZ^+Z^-,\lb{r2}\\
(Z^+)'=3\beta(X+Y)^2Z^++6\beta^2X(X+Y)Z^++9\beta^3X^2Z^++6\beta^2(Z^+)^2Z^-\lb{r3},\\
(Z^-)'=3\beta(X+Y)^2Z^-+6\beta^2X(X+Y)Z^-+9\beta^3X^2Z^-+6\beta^2Z^+(Z^-)^2\lb{r4},
\eng
where we wrote $X_{n-1}$, $Y_{n-1}$, and $Z^\pm_{n-1}$ as $X$, $Y$, and $Z^\pm$, and $X_{n}$, $Y_{n}$, and $Z^\pm_{n}$ as $X'$, $Y'$, and $(Z^\pm)'$.
See Fig.~\ref{f:XYZ}.
Since \rlb{r3} and \rlb{r4} are symmetric under the exchange of $Z^+$ and $Z^-$ and we have $Z^+_0=Z^-_0$, we see that $Z^+_n=Z^-_n$ for all $n$.

Then the relations \rlb{r1}, \rlb{r2}, \rlb{r3}, and \rlb{r4} lead to the following recursion relations for $y_n$ and $z_n=Z^\pm_n/X_n$.
\eqg
y'=\frac{3\beta^2(1+y)^2+6\beta^3(1+y)+9\beta^4+6\beta^2(1+y)z^2+18\beta^3z^2}{(1+y)^3+3\beta^2(1+y)+2\beta^3+6\beta(1+y)z^2+6\beta^2z^2},\lb{RR1}\\
z'=\frac{3\beta(1+y)^2z+6\beta^2(1+y)z+9\beta^3z+6\beta^2z^3}{(1+y)^3+3\beta^2(1+y)+2\beta^3+6\beta(1+y)z^2+6\beta^2z^2},\lb{RR2}
\eng
where we wrote $y_{n-1}$ and $z_{n-1}$ as $y$ and $z$, and $y_{n}$ and $z_{n}$ as $y'$ and $z'$.
Our task is to start from the initial values $(y_0,z_0)=(0,1)$, repeatedly apply the recursion relations \rlb{RR1} and \rlb{RR2}, and find the behavior of $(y_n,z_n)$ in the limit $n\up\infty$.
We get a reliable conclusion from a simple numerical calculation.  
We see that there is a critical value of $\beta$, which is estimated to be $\betac\simeq0.319$.
For $\beta\in(0,\betac)$, we see that $y_n\to y^*(\beta)>0$, and $z_n\to0$.
This means that the order parameter $\bktgs{\ha_o}^+$ tends to zero as $n\up\infty$, indicating that there is no BEC.
For $\beta\in(\betac,1/2)$, we see that $y_n\to y^*(\beta)>0$, and $z_n\to z^*(\beta)>0$.
Thus the order parameter $\bktgs{\ha_o}^+$ converges to a nonzero value as $n\up\infty$.
This means that the ground state exhibits spontaneous symmetry breaking of the U(1) symmetry, which corresponds to BEC.

\section{Exact energy eigenstates}
\label{A:ES}

We construct a series of exact energy eigenstate for any particle number, including $N=|\calI|$.
Theses energy eigenstates are interesting by themselves since they provide novel examples of quantum many-body scars \cite{scar1,scar2}.

Here we assume that the lattice $(\calE,\calB)$ is connected and bipartite.
By  $(\calE,\calB)$ is bipartite we mean that there is a decomposition $\calE=\calE_+\cup\calE_-$ such that two sites $x$, $y$ may be connected by a bond in $\calB$ only when $x\in\calE_+$, $y\in\calE_-$ or $x\in\calE_-$, $y\in\calE_+$.
For simplicity we shall assume that a pair of sites $x,y\in\calE$ is either connected by $p$ bonds in $\calB$ or not connected at all.
But the following result about exact energy eigenstates is valid without this restriction.

As in section~\ref{s:OPN}, we define
\eq
\hD:=\sum_{x\in\calE_+}\ha_x-\sum_{x\in\calE_-}\ha_x.
\en
Then we shall prove, for any subset $S\subset\calI$ and $n=1,2,\ldots$, that the state
\eq
\ket{\Phi_{S,n}}=(\hDD)^n(\prod_{u\in S}\hbd_u)\vac.
\lb{Phin}
\en
satisfies
\eq
\hH\ket{\Phi_{S,n}}=nt\zeta^2\ket{\Phi_{S,n}}.
\en
Thus $\ket{\Phi_{S,n}}$ is an exact energy eigenstate with particle number $N=|S|+n$ and energy $E_{S,n}=nt\zeta^2$.

Note that we can express $\hD$ in terms of the $\hd$-operators as
\eq
\hD=\frac{1}{\zeta}\Bigl(
\sum_{x\in\calE_+}\hd_x-\sum_{x\in\calE_-}\hd_x
\Bigr).
\en
Then, by using \rlb{Hhd}, we find
\eqa
[\Hhop,\hDD]&=\frac{1}{\zeta}\Bigl(
\sum_{x\in\calE_+}\Bigl\{t(\zeta^2+pz_x)\hdd_x+pt\sum_{y\in\calE(x)}\hdd_y\Bigr\}
-\sum_{x\in\calE_-}\Bigl\{t(\zeta^2+pz_x)\hdd_x+pt\sum_{y\in\calE(x)}\hdd_y\Bigr\}
\Bigr)
\nl&=\frac{1}{\zeta}\Bigl(
\sum_{x\in\calE_+}t\zeta^2\hdd_x
-\sum_{x\in\calE_-}t\zeta^2\hdd_x
\Bigr)
\nl&=t\zeta^2\hDD.
\lb{Hhd2}
\ena
We note in passing that $t\zeta^2$ is the minimum energy among single-particle energy eigenstates that are orthogonal to the flat band (i.e., the states generated by the $\hbd$ operators), and that the state generated by $\hDD$ is the unique energy eigenstate with energy $t\zeta^2$.
This fact is obvious from the dispersion relation \rlb{dispersion} for the models on the hypercubic lattice, but can be proved in general.

Note that by construction the operator $\hDD$ does not contain $\had_u$ for any $u\in\calI$.
This immediately implies that
\eq
[\Hint,\hDD]=0.
\lb{Hid}
\en
It is clear from \rlb{Hhd2} and \rlb{Hid} that $\Hhop\ket{\Phi_{S,n}}=nt\zeta^2\ket{\Phi_{S,n}}$ and $\Hint\ket{\Phi_{S,n}}=0$.
We thus find that $\ket{\Phi_{S,n}}$ is an eigenstate of $\hH=\Hhop+\Hint$ with eigenvalue $E_{S,n}=nt\zeta^2$.

\end{widetext}

%%%%%%%%%%%%%%%%%%%%%%%%

\end{document}